# Redesigning the Jefferson Lab Hall A Beam Line for High Precision Parity Experiments

Jay Benesch and Yves Roblin (7 November 2021 version)
Thomas Jefferson National Accelerator Facility

**Abstract**

The Continuous Electron Beam Accelerator Facility [CEBAF] was built with a thermionic electron source and the three original experimental hall lines reflected this. A few years after beam delivery began a parity violation experiment was approved and two polarimeters were installed in the Hall A beam line. The beam raster system was placed after the new Compton polarimeter, before one accelerator quadrupole and four quadrupoles in the new Moller polarimeter. It was very difficult to meet experimental requirements on envelope functions and raster shape with this arrangement so a sixth quadrupole was installed downstream of the Moller polarimeter to provide an additional degree of freedom. All of the parity experiments in Hall A have been run with this still-unsatisfactory configuration. The MOLLER experiment is predicated on achieving a 2% error on a 32 ppb asymmetry. Beam line changes are required to meet the systematic error budget. This paper documents the existing beam line, an interim change which can be accomplished during a annual maintenance down, and the final configuration for MOLLER and subsequent experiments.

**Parity Experiment requirements**

The history of parity violation (PV) experiments is summarized in references [3] and [4]. The first CEBAF PV experiment was HAPPEX [5]. Table 1 below is taken from [6], the submission to PRL of the latest JLab parity experiment to complete analysis, PREX-II.

**Table 1**. Corrections and systematic uncertainties to extract $A_{meas}$ PV listed on the bottom row with its statistical uncertainty.

| Correction | Absolute (ppb) | Relative (%) |
|---|---|---|
| Beam asymmetry | −60.4 ± 3.0 | 11.0 ± 0.5 |
| Charge correction | 20.7 ± 0.2 | 3.8 ± 0.0 |
| Beam Polarization | 56.8 ± 5.2 | 10.3 ± 1.0 |
| Target diamond foils | 0.7 ± 1.4 | 0.1 ± 0.3 |
| Spectrometer rescattering | 0.0 ± 0.1 | 0.0 ± 0.0 |
| Inelastic contributions | 0.0 ± 0.1 | 0.0 ± 0.0 |
| Transverse asymmetry | 0.0 ± 0.3 | 0.0 ± 0.1 |
| Detector nonlinearity | 0.0 ± 2.7 | 0.0 ± 0.5 |
| Angle determination | 0.0 ± 3.5 | 0.0 ± 0.6 |
| Acceptance function | 0.0 ± 2.9 | 0.0 ± 0.5 |
| Total correction | 17.7 ± 8.2 | 3.2 ± 1.5 |
| $A_{meas}$ PV and statistical error | 550 ± 16 | 100.0 ± 2.9 |

The Qweak parity experiment final results [6]: Asymmetry-226.5 ±7.3 ppb (stat) ±5.8 ppb (syst); total uncertainty 9.3 ppb or 4.1%. Systematic error was 2.6% so even if beam time had been doubled to lower the statistical error, the total error would have been 3.4%. This is the smallest asymmetry yet measured at CEBAF. The adjustment to the measured result was not couched in the same terms used in Table 1, see the paper. The NIM report on the Qweak apparatus may also be of interest. [15] Per the MOLLER final conceptual design report [8], the expected experimental asymmetry is ~32 ppb.

**Table 2: Expected fractional errors** are (Table 3 of CDR, Ref 8, page 22 of pdf)

| Error Source | Fractional Error [%] |
|---|---|
| Statistical | 2.1 |
| Absolute Norm. of the Kinematic Factor | 0.5 |
| Beam [second order] | 0.4 |
| Beam polarization | 0.4 |
| e + p[+γ] → e + X[+γ] | 0.4 |
| Beam [position, angle, energy] | 0.4 |
| Beam [intensity] | 0.3 |
| e + p[+γ] → e + p[+γ] | 0.3 |
| γ[∗] + p → [π,μ,K] + X | 0.3 |
| Transverse polarization | 0.2 |
| Neutral background [soft photons, neutrons] | 0.1 |
| Linearity | 0.1 |
| Total systematic | 1.1 |

The fractional systematic error required by MOLLER is 40% of that achieved by Qweak and the absolute systematic error needed is 0.35 ppb versus 5.8 ppb for Qweak, a factor of sixteen. The CEBAF injector is being rebuilt with a higher kinetic energy source [less space charge effect] and focusing elements with one tenth the focusing variation across the full beam width (6σ) than was the case for the experiments performed to date. This is done to reduce helicity correlated beam parameters so fast and slow reversals will cancel asymmetries. It must be possible to measure most of the parameters in Table 2 in the Injector and Hall A and feed back as possible to minimize their effects integrated over the course of the expected four years (100 weeks beam time) of operation. The Hall A beam line must also be rebuilt and diagnostics added to make possible (if still unlikely) the required fractional errors. The 2020 MOLLER Conceptual Design Report [8] ties these requirements to hardware implementation. The evolution of the CEBAF injector is dealt with in [9]. The remainder of this paper will address the changes in the Hall A beam line. An element list including magnets, diagnostics, drifts and some vacuum equipment follows. As is standard in accelerator physics codes, diagnostics and steering magnets are represented as points. Drifts, dipoles, quadrupoles and raster magnets are the only elements with finite length.

**Present beam line**

Beam envelopes with design Twiss parameters and as-measured emittances at 11 GeV are shown in Figure 1. Full vertical scale is 1 mm; minimum beam pipe ID is 22 mm. All optics work by the first author is done in OptimX [1] and most figures are generated therein.

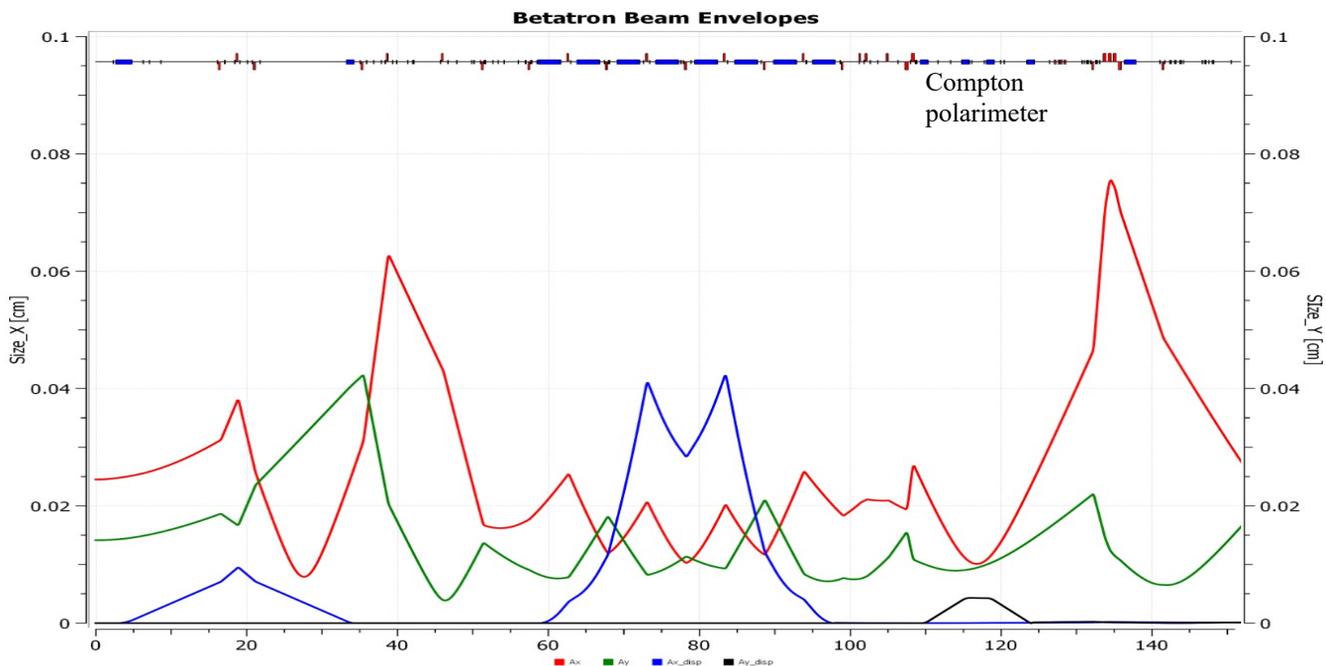

**Figure 1**. Normal target location is at the right edge of the plot. The horizontal beam size (red) is almost twice the vertical (green). This is a result of the competing constraints on the beam with inadequate variables. The blue line is the horizontal beam size due to dispersion with dp/p 2E-4. The black line is the size due to vertical dispersion in the Compton polarimeter chicane. These need to be summed in quadrature with the red and green values, respectively. In the line at the top of all the images, red indicates quadrupoles and blue dipoles. X axis m.

A 5 mm square raster at the target is standard. The figure below shows the results with the horizontal raster [red] at power supply maximum 50 A and the vertical at 34 A. The raster is driven by a triangle wave at 25 kHz; it suffices to show DC result for those currents. The requirement that 2.5 mm half-response be obtained at 50 A prevents the horizontal and vertical envelopes from being matched in Figure 1 given six quadrupoles between raster and target. .

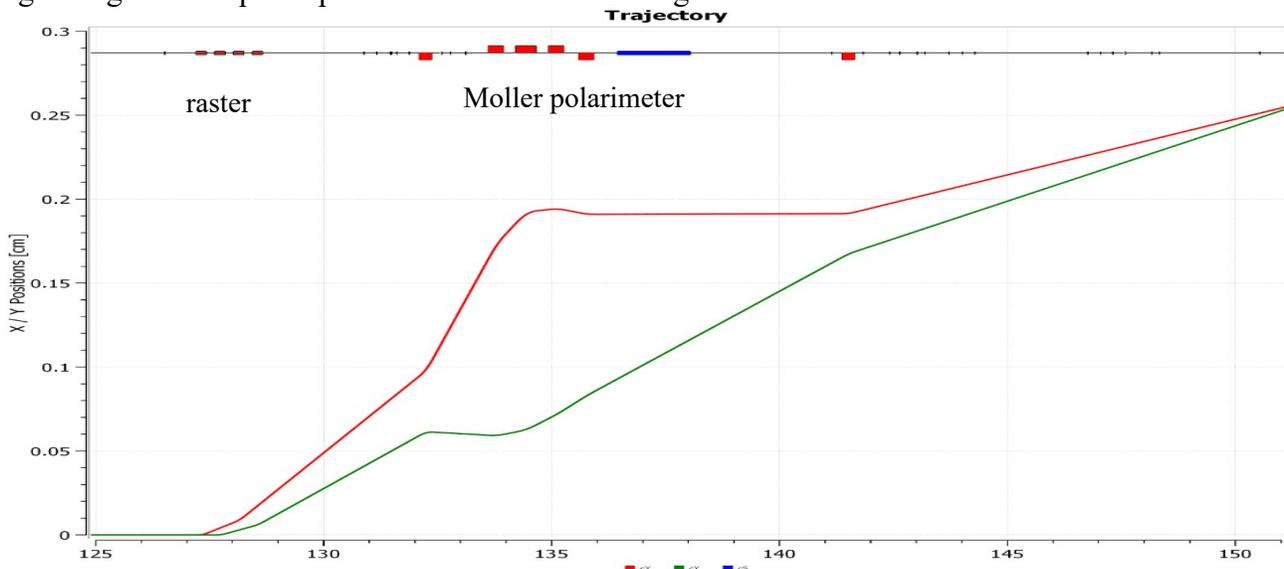

**Figure 2**. Raster response with subsequent quadrupoles set to provide envelopes in Figure 1. 50 A horizontal and 34 A vertical drive. Note the horizontal scale change from figure 1; this figure shows only the last 26 m to the pivot vs 151 m there. X axis meters. Right axis unused.

In Figure 2 one sees the four quads of the Moller polarimeter [red] are quite close together and so do not act fully independently as they would with larger drifts between them.  Blue is the polarimeter dipole which is energized only when polarization is being measured.  Among the requests which are not fulfilled by this optics is the preference for a round spot on the Moller polarimeter iron target foil; the ratio shown in figure 1 is 4:1 x:y.  Only the quadrupoles downstream of the Compton polarimeter (four dipoles at S=100-120 m in Figure 1) may be used for final focus and raster adjustment as those before must be used to achieve a 100 micron round spot where the polarized laser beam intersects the electron beam.  It is also desirable to have zero derivatives of the envelope [beta] functions at the target so small changes in input betas and alphas do not affect the beam size at the target.  One needs to maintain an envelope less than a mm throughout the beam line, even if the input emittances are larger than specified, in order to reduce the generation of beam halo which will become background in the nuclear physics spectrometers.  There are effectively four variables available via the six quadrupoles in Figure 2 yet ten parameters which the beam line designer would like to minimize.

CEBAF accelerator physics and operations personnel have contended with this for two decades.  The first author began designing alternatives a decade ago but there was no compelling reason to make changes until the MOLLER [2] experiment was funded.  The need for almost an order of magnitude improvement in systematic error over Qweak finally mandated change.  The final optics for that experiment, which will remain in Hall A thereafter and be used for all subsequent experiments, is described in what follows.

**Basic constraints**

The principal constraints are the spatial length of the MOLLER experiment and the diameter of Hall A.  In 2009, when the first author first designed an altered beam line for Hall A in concert with the first MOLLER proposal to the Jefferson Lab Physics Advisory Committee, the center of the LH2 target was to be 6 m upstream of the center of the hall, the normal pivot.  The only way to meet all the constraints was to use the Moller polarimeter quads as part of the production beam focusing elements.  Since it was also necessary to move the raster downstream of all focusing elements, this required moving the entire polarimeter upstream.

In 2018 a detailed cost analysis of the beamline rework was done.  The cost to move the polarimeter upstream was sufficiently large that management required it be kept in place.  The MOLLER experiment target and detector system had evolved by this time so the target center was now 4.5 m from the center of the hall rather than 6 m.  This additional beamline space downstream of the polarimeter for diagnostics and the upstream space made available by fixing the polarimeter allowed the production beam transport system to change.  The six polarimeter magnets would be energized only during polarization measurements and degaussed thereafter.  A triplet of 2.85 cm ID accelerator magnets with spacing about 220 cm would be used to re-focus the round beam coming out of the Compton polarimeter.  The raster would follow the accelerator quads; its response through the degaussed polarimeter would be the same as through a drift.

One sees (blue line) in Figure 1 that the nominal 12 GeV Hall A arc optics has two dispersion peaks at x ~65 m and x ~95 m; the dispersion is about 1.5 m there.  This was done to reduce the energy spread generated by synchrotron radiation in the arc.  The collaboration requests that there be one dispersion peak of 4 m, as in all previous parity experiments, to improve resolution of the on-line energy monitor (Figure 3).  Accordingly, the strengths of the quads upstream of the reworked region will be changed.  No elements will be moved, only the current supplied will change.  The resulting optics will be shown in a later section.

**Element list**

N is the element number within the entire Hall A beam line. Name is that used by the EPICS control system. S is the distance along the beam path from the start of the Hall A beam line to the end of the specified element. L is the element length, steel-only for magnets, not effective length. BCM is a cavity-based Beam Current Monitor. Unser is different type of current monitor, used to calibrate the others. This was also generated in OptimX [1].

| N | Name | description | S[cm] | L[cm] | G[kG/cm] |
|---|---|---|---|---|---|
| 235 | GMCP1P04 | end of Compton polarimeter | 12453.5 | 0 | |
| 236 | oD4000 | drift | 12470.8 | 17.27 | |
| 237 | iIHV1H00 | gate valve | 12470.8 | 0 | 0 |
| 238 | oD4001 | drift | 12480.5 | 9.68 | |
| 239 | iIPM1H01 | short stripline BPM center | 12480.5 | 0 | 0 |
| 240 | oD4002 | drift | 12495.5 | 15.02 | |
| 241 | qMQK1H01 | quadrupole | 12526 | 30.48 | 2.63 |
| 242 | oD4003 | drift | 12545.1 | 19.08 | |
| 243 | kMCG1H01H | horizontal corrector center | 12545.1 | 1.00E-06 | 0 |
| 244 | oD4004 | drift | 12564.7 | 19.61 | |
| 245 | kMCG1H01V | vertical corrector center | 12564.7 | 1.00E-06 | 0 |
| 246 | oD5000 | drift | 12583.6 | 18.95 | |
| 247 | iITV1H01 | viewer | 12583.6 | 0 | 0 |
| 248 | oD5001 | drift | 12698.8 | 115.21 | |
| 249 | iIPM1H02 | short stripline BPM center | 12698.8 | 0 | 0 |
| 250 | oD4009 | drift | 12711.3 | 12.48 | |
| 251 | qMQR1H02 | quadrupole | 12746.9 | 35.56 | -3.4 |
| 252 | oD4010 | drift | 12762.9 | 16.01 | |
| 253 | kMCG1H02H | horizontal corrector center | 12762.9 | 1.00E-06 | 0 |
| 254 | oD4011 | drift | 12782.5 | 19.61 | |
| 255 | kMCG1H02V | vertical corrector center | 12782.5 | 1.00E-06 | 0 |
| 256 | oD5002 | drift | 12816.2 | 33.67 | |
| 257 | iIBC1H02A | Cavity Beam Current Monitor BCM | 12816.2 | 0 | 0 |
| 258 | oD5003 | drift | 12836.2 | 20.02 | |
| 259 | iIBC1H02B | BCM | 12836.2 | 0 | 0 |
| 260 | oD5004 | drift | 12877.4 | 41.18 | |
| 261 | iIUN1H02 | Unser Current Monitor | 12877.4 | 0 | 0 |
| 262 | oD5005 | drift | 12917.2 | 39.82 | |
| 263 | iIPM1H03 | short stripline BPM center | 12917.2 | 0 | 0 |
| 264 | od4013 | drift | 12932.2 | 15.02 | |
| 265 | qMQK1H03 | quadrupole | 12962.7 | 30.48 | 2.27 |
| 266 | oD4014 | drift | 12981.8 | 19.08 | |
| 267 | kMCG1H03H | horizontal corrector center | 12981.8 | 1.00E-06 | 0 |
| 268 | oD4015 | drift | 13001.4 | 19.61 | |
| 269 | kMCG1H03V | vertical corrector center | 13001.4 | 1.00E-06 | 0 |
| 270 | oD4016 | drift | 13033 | 31.63 | |
| 271 | krastX | horizontal raster coil | 13058 | 25 | 0 |
| 272 | oD4017 | drift | 13068.6 | 10.56 | |
| 273 | krastY | vertical raster coil | 13093.6 | 25 | 0 |
| 274 | oD5006 | drift | 13113.7 | 20.13 | |

| N | Name | description | S[cm] | L[cm] | G[kG/cm] |
|---|---|---|---|---|---|
| 275 | iIPM1H04 | short stripline BPM center | 13113.7 | 0 | 0 |
| 276 | oD4019 | drift | 13125.4 | 11.74 | |
| 277 | iIHA1H04 | wire scanner | 13125.4 | 0 | 0 |
| 278 | oD4020 | drift | 13135.3 | 9.85 | |
| 279 | iIHV1H04 | gate valve | 13135.3 | 0 | 0 |
| 280 | oD5007 | drift | 13161.1 | 25.81 | |
| 281 | iIPM1H04AX | nA cavity BPM X | 13161.1 | 0 | 0 |
| 282 | oD5008 | drift | 13190.6 | 29.52 | |
| 283 | iIPM1H04AY | nA cavity BPM Y | 13190.6 | 0 | 0 |
| 284 | oD4023 | drift | 13218.4 | 27.76 | |
| 285 | iIBC1H04A | cavity BCM | 13218.4 | 0 | 0 |
| 286 | oD5009 | drift | 13244.3 | 25.91 | |
| 287 | iIPM1H05 | short stripline BPM center | 13244.3 | 0 | 0 |
| 288 | oD5010 | drift | 13281.2 | 36.97 | |
| 289 | iIMollTar | Moller polarimeter target foil | 13281.2 | 0 | 0 |
| 290 | oD4026 | drift | 13360.8 | 79.6 | |
| 291 | qMQO1H06 | Moller polarimeter quadrupole 1 | 13397.1 | 36.22 | 0 |
| 292 | oD4027 | drift | 13421.8 | 24.71 | |
| 293 | qMQM1H07 | Moller polarimeter quadrupole 2 | 13466.8 | 45.05 | 0 |
| 294 | oD4028 | drift | 13494 | 27.2 | |
| 295 | qMQO1H08 | Moller polarimeter quadrupole 3 | 13530.2 | 36.22 | 0 |
| 296 | oD4029 | drift | 13560 | 29.78 | |
| 297 | qMQO1H09 | Moller polarimeter quadrupole 4 | 13596.2 | 36.22 | 0 |
| 298 | oD4030 | drift | 13642 | 45.71 | |
| 299 | bMMA1H10 | Moller polarimeter dipole | 13803.8 | 161.8 | 0 |
| 300 | oD5011 | drift | 14128.2 | 324.41 | |
| 301 | iIPM1H11 | 20 cm BPM center | 14128.2 | 0 | 0 |
| 302 | oD5012 | drift | 14144.6 | 16.41 | |
| 303 | iIHV1H11 | vacuum valve | 14144.6 | 0 | 0 |
| 304 | oD4033 | drift | 14152.6 | 8 | |
| 305 | iIHA1H11 | wire scanner | 14152.6 | 0 | 0 |
| 306 | oD5013 | drift | 14188.4 | 35.85 | |
| 307 | iIBC1H12A | electrically isolated cavity BCM | 14188.4 | 0 | 0 |
| 308 | oD5014 | drift | 14215.8 | 27.41 | |
| 309 | iIBC1H12B | electrically isolated cavity BCM | 14215.8 | 0 | 0 |
| 310 | oD5015 | drift | 14243.7 | 27.87 | |
| 311 | iIBC1H12C | electrically isolated cavity BCM | 14243.7 | 0 | 0 |
| 312 | oD5016 | drift | 14284.3 | 40.56 | |
| 313 | iIPM1H13AX | nA cavity BPM X | 14284.3 | 0 | 0 |
| 314 | oD4038 | drift | 14313.8 | 29.52 | |
| 315 | iIPM1H13AY | nA cavity BPM Y | 14313.8 | 0 | 0 |
| 316 | oD5017 | drift | 14341.5 | 27.76 | |
| 317 | iIBC1H13 | cavity BCM | 14341.5 | 0 | 0 |
| 318 | oD5018 | drift | 14376.8 | 35.26 | |
| 319 | iIHA1H14 | wire scanner | 14376.8 | 0 | 0 |
| 320 | oD4041 | drift | 14393.2 | 16.41 | |
| 321 | iIPM1H14 | 20 cm BPM center | 14393.2 | 0 | 0 |
| 322 | oD5019 | drift | 14425.9 | 32.66 | |

| N | Name | description | S[cm] | L[cm] | G[kG/cm] |
|---|---|---|---|---|---|
| 323 | iIBC1H15 | MPS BCM | 14425.9 | 0 | 0 |
| 324 | oD4043 | drift | 14439.5 | 13.61 | |
| 325 | iIHV1H15 | gate valve | 14439.5 | 0 | 0 |
| 326 | oD5020 | drift | 14605.1 | 165.6 | |
| 327 | iIMOLLER | MOLLER LH2 target center | 14605.1 | 0 | 0 |
| 328 | oD4045 | drift | 15055.1 | 450 | |
| 329 | iPivot | pivot in center of hall | 15055.1 | 0 | 0 |
| 330 | oD3025a | drift | 15555.1 | 500 | |
| 331 | iSolid | future SoLID target location | 15555.1 | 0 | 0 |
| 332 | oD3025b | drift | 17305.1 | 1750 | |
| 333 | iDetPlne | MOLLER detector plane | 17305.1 | 0 | 0 |
| 334 | oD3026 | drift | 17705.1 | 400 | |
| 335 | iHLwall | wall of Hall A | 17705.1 | 0 | 0 |
| 336 | oD3027 | drift | 19905.1 | 2200 | |
| 337 | iDumpFc | dump face | 19905.1 | 0 | 0 |

**Discussion**

During all beam transport except Moller polarimetry, quadrupoles 241, 251 and 265 form a triplet with 219 cm center to center spacing. The central quad is longer and has a different pole shape to provide greater focusing range. Correctors are small steering magnets with ~30 kG-cm capability, bipolar.

BCM is a cavity-based Beam Current Monitor. Unser is a different type of current monitor [10]. Seven Physics BCMs are present in the beam line, two associated with the Unser [257, 259, and Unser 261]; two associated with cavity position monitors [285, 317] and three in a separate enclosure [307, 309, 311]. The last three are electrically isolated from each other and the beam line so the only ground is through the signal cable. The distance between these cavities will provide ~160 dB of cavity to cavity RF signal isolation even without the ceramic break. It is intended that these be the primary current monitors and expected that the error allowance on beam intensity in Table 2 will be met. The BCMs had a noise floor in Qweak ~2.5 ppb [7]. An attempt to remove this floor is [16]. The BCM/BCM pair downstream of the second girder is isolated from ground and the two are at the same potential. The Unser is not isolated from beam line ground except by its construction. The seven BCMs are calibrated with beam against the Unser multiple times each run. The eighth BCM, 323, is used in the Machine Protection System (MPS). There is one in each of the halls. The sum of the hall MPS BCMs is subtracted from a similar unit at the end of the injector. If the difference is greater than ~1% the MPS system turns off beam at the source and the Operations group works to find the location and cause of the beam loss. The 1% allowance is needed in part because the MPS BCMs are not calibrated with beam against the Unser systems in Halls A and C; there are no Unsers in the Halls B and D lines as their currents are too low to accurately register. The MPS BCMs are tuned to resonance at a fixed temperature (41 C) and maintained there with insulated and heated jackets via a per-BCM temperature feedback system.

Position monitors are four-antenna stripline devices. There are two types in the proposed Hall A line, one with wire antennas (20 cm, present standard) and a shorter version with machined antennas (14 cm, elements 239, 249, 263, 275, 287). The latter is needed to make everything fit. The short machined versions have only been used at low energy in a test beam. The RF engineer responsible for the design states that it should have the same resolution as the older wire design down to 10 nA. For lower current, used in tracking studies to determine $Q^2$, position monitors consisting of two cavities sensitive to off-axis position and one to current are used, elements 281/283/285 and 313/315/317. The current signal is used to normalize the output of the position cavities. Position monitors 275 and 321 are 1269 cm apart. The collaboration specified a distance of at least 1000 cm to allow sufficient beam angle resolution. Cavities monitoring low current X positions, 281 and 313, are 1115 cm apart, also exceeding the requirement.

Wire scanners are an invasive diagnostic used to measure beam size via a slow scan of a 25 or 50 micron wire across a low to moderate (25 uA) beam, elements 277, 305 and 319. Three wire scanners allow one to fit a parabola to the X and Y beam sizes to determine beam size at the target center or another location of interest, for instance the Moller polarimeter target.

The Moller polarimeter is comprised of elements 289 through 299. Position monitor 301 is mounted above the downstream end of the polarimeter detector box. Since the latter is at a lower height it is not in the element list. Four quadrupoles are used to deflect and focus the half-energy Moller-scattered electrons into tubes beside the main beam line; they are then deflected downward into the polarimeter detector by the dipole. These elements are shown as having zero field/gradient because they aren't used during production beam transport and will be demagnetized after each measurement. Space has been left in the beam line for air core, iron return correctors atop the polarimeter detector box. These would provide steering in case beam deflections in the polarimeter quads and dipole during polarimetry are sufficient to move the unscattered beam to an unfortunate location. Conceptual design and preliminary engineering is complete. If installed, they would also be degaussed after each polarization measurement. The magnetized iron foil target of the polarimeter is moved 30 cm upstream from present location in order to reduce the Levchuk effect [12] and allow the error to be reduced to the 0.4% needed for Table 2. The Compton polarimeter will also be upgraded; that is not covered in this paper. Transverse polarization fraction can only be measured (and nulled) in the 6 MeV/c region of the injector, using a Mott polarimeter. However, the injector is set up to change polarization from longitudinal to transverse by a well defined 90° rotation so measurement of beam normal spin asymmetries can be made in the hall to provide information on the magnitude of that helicity-correlated asymmetry and include its effect in the final result.

Helicity-correlated beam position, angle and energy variations must be measured and corrected for numerically and via feedback systems. See Table 2 for error quota. The measurement is done by equipping the position monitors with fast sample-and-hold cards so helicity-dependent signals can be measured within each helicity window. The numerical correction can be derived in two ways: analysis of response to random jitter upstream in the Hall A beam line or small driven responses. A 240 Hz oscillator is switched to one of five devices so driven data can be obtained, generally for a minute per device per hour. The energy variation is provided by small signal input to the drive in one of the South Linac cryomodules. Position variation is provided by two small horizontal and two small vertical correctors in the Hall A beam line arc. The X/Y pairs are separated by enough phase advance that the responses of the BPMs after the quad triplet, e.g. elements 275 and 321, are not degenerate and can be deconvolved. See Figure 5 for coil locations.

In Table 2 Beam (second order) effects are budgeted for an error as large as first order. This is in large part because they cannot be measured on the electron beam, only on the laser before it hits the photo-cathode.

The raster is placed after all magnetic elements which are powered during production running. The shaped response seen in figure 2 is therefore absent: the response is a straight line. See Figure 6. Only a single set of raster coils can be used on the new beam line due to the MOLLER target position 450 cm upstream of the usual location. For a target at the usual location, the single X coil requires 25% more current than each of the two X coils in Figure 2. The single Y coil requires 80% more current than each of the two Y coils in Figure 2. For a target at the usual location, the currents needed in the single raster coil of each plane are 77% of those at the MOLLER target center for same raster span. The raster frequency is 25 kHz. A test has been run with the existing raster power supplies which can provide a 60 A zero to peak triangle wave. Equilibrium temperature was measured at 10 A intervals and a parabola fit to the data. The extrapolated temperature at 80 A, the needed current for a 5 mm square raster at target entrance window, is 64 C. At 100 A the projected temperature is 84 C. The insulation on the existing coils is polyurethane with nylon overcoat, 130 C continuous rating. The test will be repeated when the requested 100 A zero to peak power supply has been prototyped.

**Steering and alignment**

The BPMs which will be used for the slow orbit lock are 1H11 and 1H14, elements 275 and 321, which are 1269 cm apart. The distance between 1H14 (321) and the MOLLER detector plane (333) is 2951.5 cm. It follows that angle steering errors are multiplied by 2.3. There are six radial detector rings with varying tile thetas. The detector ring for the elastic scattered electrons, ring 2, has 40 mm radial extent. The tail of the elastic distribution extends into the larger radius rings, including ring 5 which includes most of the Moller scattered electrons. From Figure 30 of Ref. [8] page 57 (pdf page 71), a mm shift in beam location is contraindicated if one wants to keep the elastically scattered electrons predominately in ring 2. It follows that angle error exiting the slow orbit lock BPM pair should be 15 μrad or less. Achieving this will be difficult. The random angle fluctuation at 1 kHz [helicity pair] is required to be 4.7 μrad per Table 5 of Ref. [8] page 15 (p 29 of pdf). The slow orbit lock is generally run with five seconds between updates. Given the accelerator diagnostic monitoring system update rate of 1 Hz, two second intervals are the best that can be done without substantial upgrades.

As mentioned in the discussion of the Moller polarimeter above, it may be necessary to add a pair of correctors just after the polarimeter detector to steer unscattered beam to the dump. These would be degaussed with the polarimeter so they do not affect beam during MOLLER production running. They would likely be in regular use for later experiments.

Null collimator and spectrometer transverse offsets [x,y] and angles [pitch, yaw] are difficult to achieve. If all the correctors in the element list are have 18 kG-cm strength, only 0.7 mm vertical offset can be dealt with if pitch and yaw are zero. The horizontal plane has 4 mm capability in this case, which should suffice. This corrector has been modeled in 3DS Opera. It appears that the maximum strength may be doubled via new power supplies if additional cooling is provided and more turns included. The wire is 2 mm square and should take 20 A, versus 12 A maximum now. The new corrector design is denoted MCG. Magnetic design is complete and mechanical design almost so.

The LH2 target is 125 cm long with 127 micron aluminum alloy beam windows. There are solid targets on the same vertical ladder to allow for systematic studies including measurement of the scattering from the windows on the LH2 target. The end of the target is 145.676 m from the start of the Hall A line; 147 m will be the terminus in figures 3-5 which follow.

## MOLLER optics

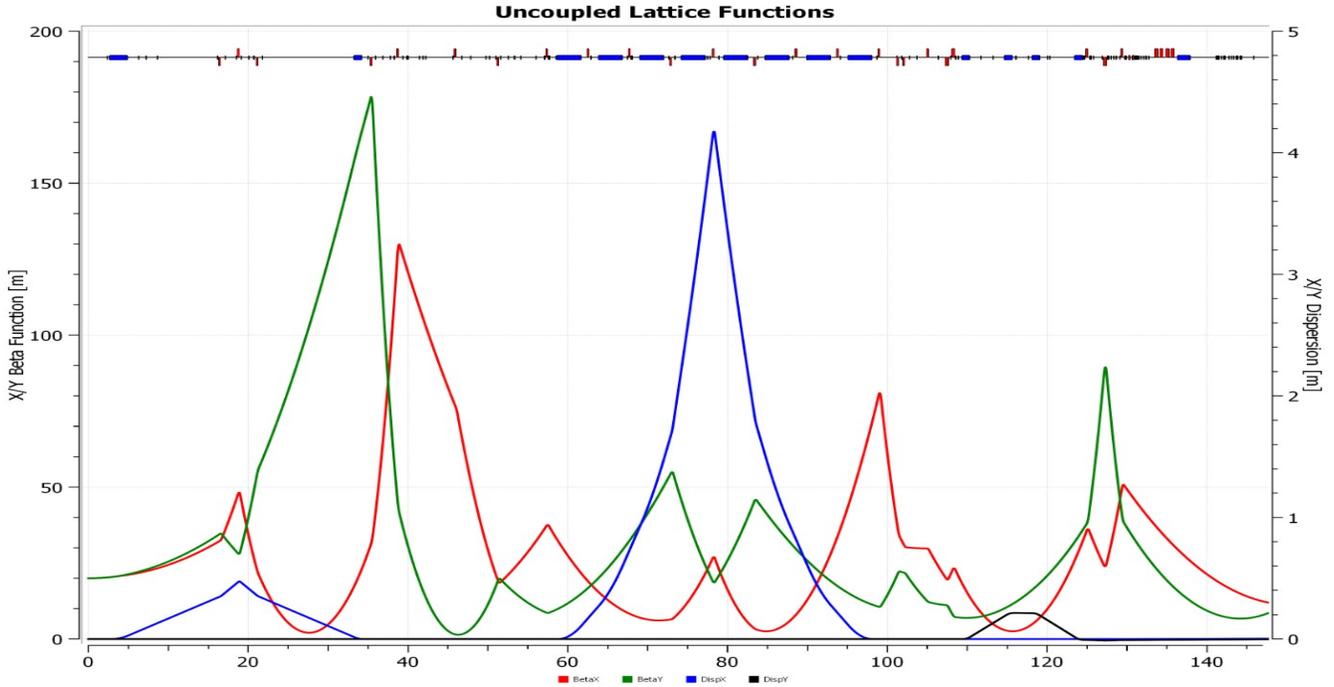

**Figure 3**. Beta functions and dispersion throughout Hall A line to 32 cm past end of LH2 target. Blue is horizontal dispersion (right axis, m) and is peaked at the center of the arc leading to Hall A.

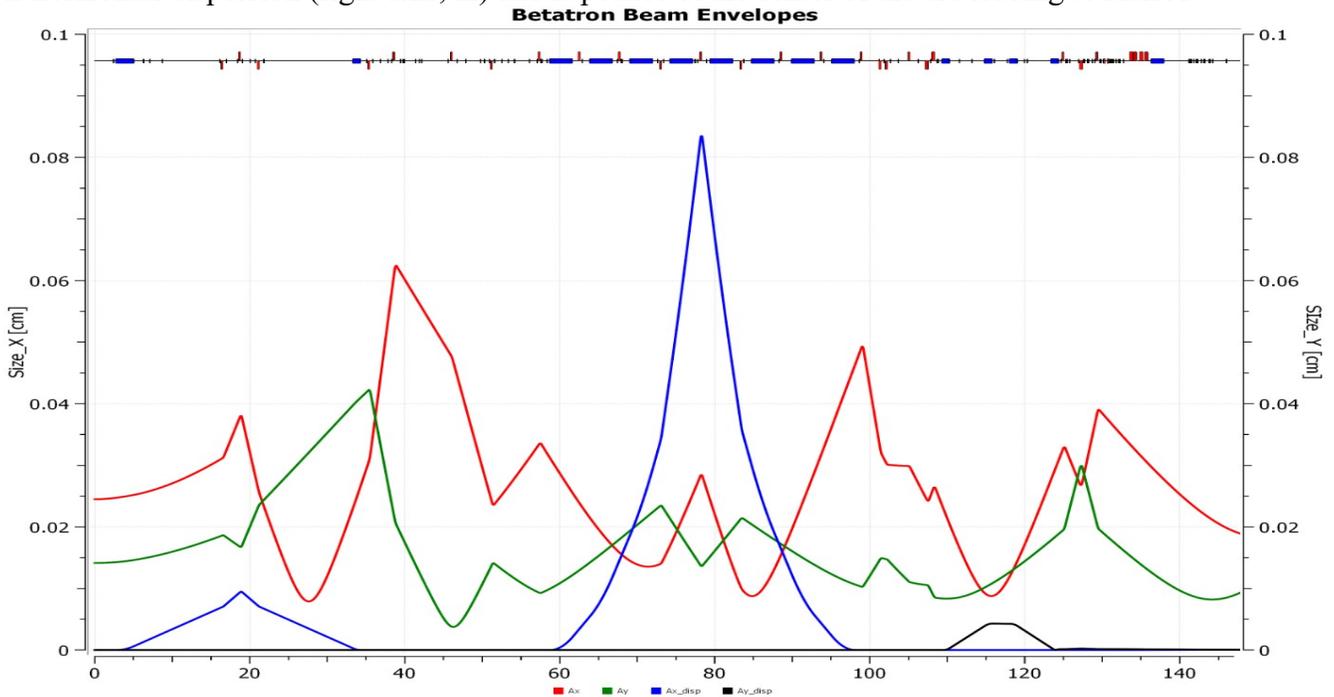

**Figure 4**. Beam envelopes, 1 mm full vertical scale, from start of Hall A transport line through to 32 cm past end of LH2 target. Horizontal (red) minima at right are at Compton polarimeter interaction point and at center of MOLLER LH2 target. Horizontal emittance is three times vertical so more focusing is needed. The quadrupole triplet, elements 241, 251 and 265, may be relaxed to move the final minimum either to the pivot or to ~4 m beyond it. Beam size would still be within agreed specification when relaxed to the latter extent. Blue line is envelope due to 4 m dispersion coupled to 2E-4 momentum spread; horizontal position at the 1C12 BPM, at the peak, is dominated by energy changes (Fig. 3) just as beam envelope is.

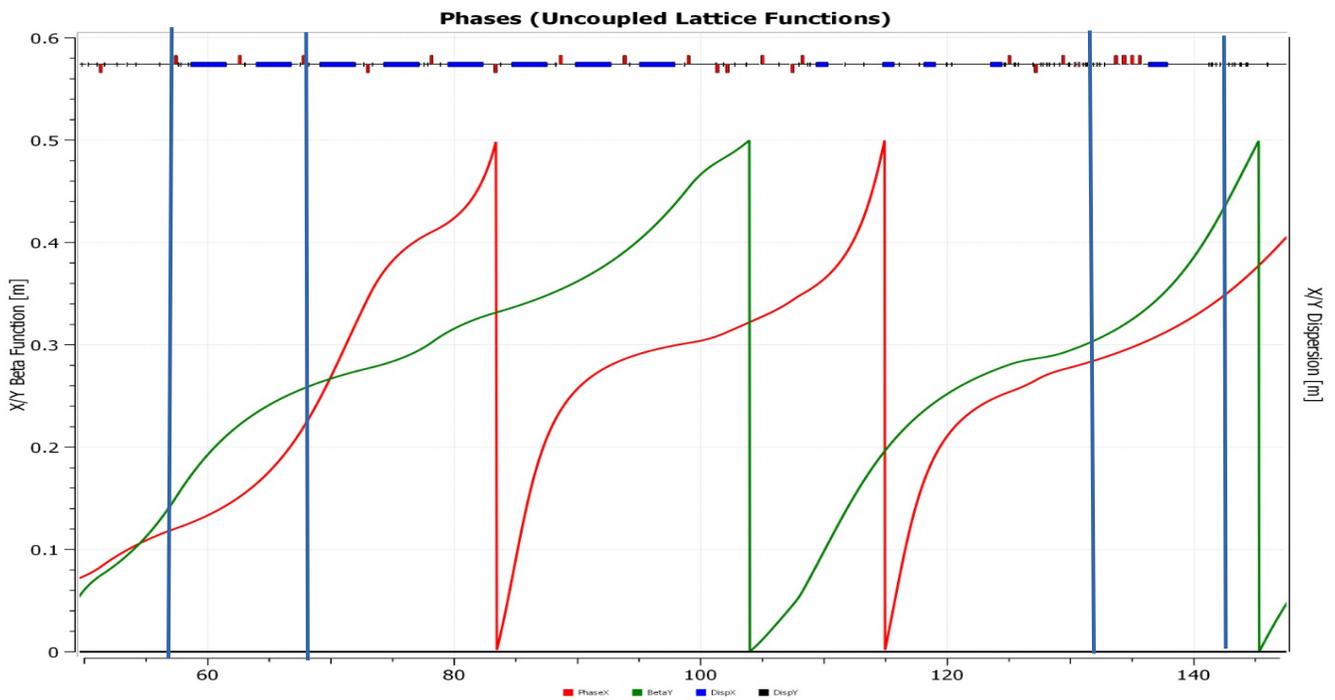

**Figure 5**. Phase advance from 1C07 quad to 32 cm past end of LH2 target. The two vertical lines at left indicate the position of H/V coils which can be used to excite small position variations to quantify helicity dependent position differences. The two lines at the right are BPMs after all excited magnets. There is adequate but not optimal phase difference between the exciting coils and the BPMs. There are two intermediate BPMs on the right not indicated. The four BPMs in question are elements 275, 287, 301 and 321. Two cavity BPMs are also in this region. If the phase advance proves inadequate one of the coil pairs may be moved farther upstream during or after the first sixteen week MOLLER run.

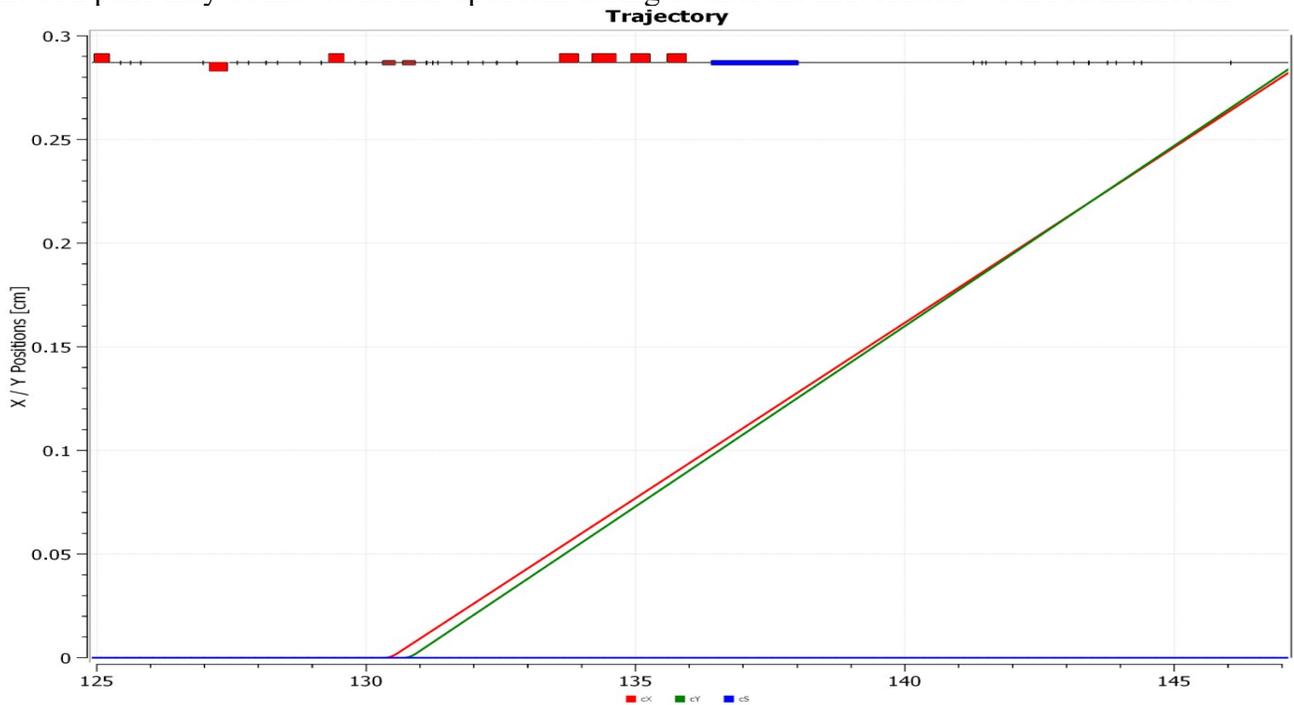

**Figure 6**. Raster response is linear as all magnets after the raster are degaussed during production running. 74 A in vertical coil, green, for 5 mm square at center of target; 80 A for 5 mm square at upstream end of target chamber.

## An interim change

The experiments tentatively scheduled for May-December 2022 use a polarized He3 target in a glass cell 600 mm long by 19 mm Inner Diameter (ID). Target cells are described in [13]. These are handmade by a university glassblower and are not uniform in dimension. As a result the target cell may be off the beam line axis by a few mm at either end, creating both position and angle offsets. The present beam line, Figures 1 and 2, has inadequate capacity for beam position and angle adjustment. It has been proposed that a portion of the MOLLER beam line rework be done January-April 2022. The third quadrupole girder, elements 262-269, would be replaced by the second existing raster girder so existing raster power supplies can be used. The nA cavity BPM assembly, elements 280-286, would remain in its current location on a long diagnostic girder between the Moller polarimeter and the Pivot at the center of the hall. Changes associated with elements 300-328 would not be made at this time. The doublet, elements 241 and 251, does not suffice to create a minimum at the pivot so the Moller polarimeter quads are pressed into service; raster distortion is still much less than shown in Figure 2.

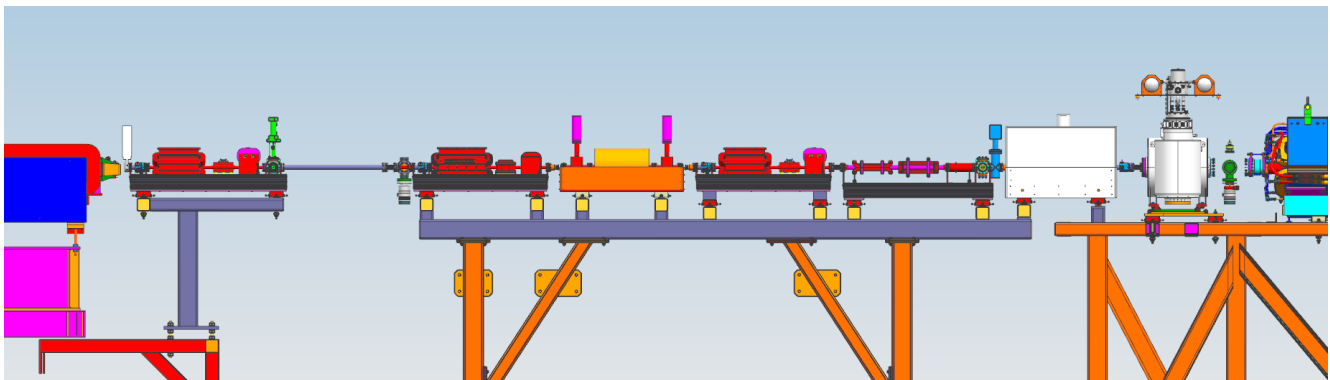

**Figure 7**. First part MOLLER beam line as described above.

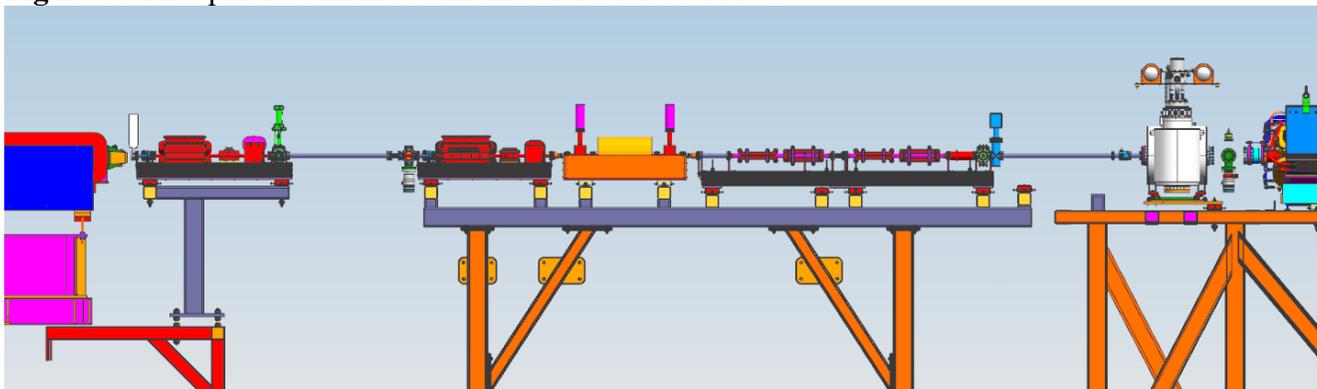

**Figure 8**. Proposed interim beam line. Third accelerator girder is replaced with existing second set of raster coils. nA BPM [white box] is left on diagnostic girder. Moller polarimeter target [white cylinder at right] is moved upstream 30 cm as requested to increase focusing to reduce the Levchuk effect. The last dipole in the Compton polarimeter chicane is at left.

This interim arrangement may be optimized in two manners: one may minimize the spot envelope at the Pivot, at the cost of requiring very different currents in the X (41 A) and Y (25 A) rasters to get equal deflection at the pivot, or one may optimize the raster and create a larger spot. In the latter case the currents are 30 A in X and 28 A in Y. This makes it trivial to get a nearly circular raster as desired for the glass target. The existing raster power supplies are rated at 60 A but limited to 50 A for improved reliability. For MOLLER new power supplies are required as the single raster coil per plane requires 80 A.

There is a major improvement in steering capability with the modified design. The existing deck can only manage up to 1.9mm target offset if wanting to come with zero angle or 3.7 mm if not requiring a zero angle. The interim design has a very wide range of X target offset it can accommodate, over 8mm, while also being able to come in with zero angle which is a major advantage for the He3 long target.

In both instances it is assumed one starts with correctors at zero. Obviously if they were not the range may be less. While this is not a problem for the interim design since there is so much headroom, it's an issue with the existing optics. This is what we have observed over two decades. Around 3mm target offset correction is difficult. Sometimes we rail correctors and have to make suboptimal changes in orbit upstream, compromising the orbit in the Compton polarimeter, in order to get through the offset/tilted target and on to the dump.

With the interim solution, one can implement a "smart knob", aka a simultaneous software control for multiple correctors. The recipe is: Power MCG1H01H and then set MCG1H02H 60% weaker, set MCG1H04H like 1H02H but with sign reversed. MCG1H04H is on a girder after element 299 in the list above; this girder will be removed in the final beam line.

Figure 9 below shows correction for 1mm target offset in X/Y with the interim solution and correctors H01/H02/H04. It was calculated in ELEGANT and the solution typed into Optim to generate the plot. There are also solutions for which one can decide to use H01/H04 or H02/H04 instead of all three. For MOLLER the three X/Y pairs of correctors (243/245, 253/255, 267/269) will suffice; the 1H04 corrector pair will not exist.

This interim solution may also be used after the completion of the MOLLER experiment as the two raster girders will allow a 10 mm radius raster on a polarized NH3 target placed a few meters after the pivot for SIDIS experiments using the SoLID apparatus if the latter is funded. [14]

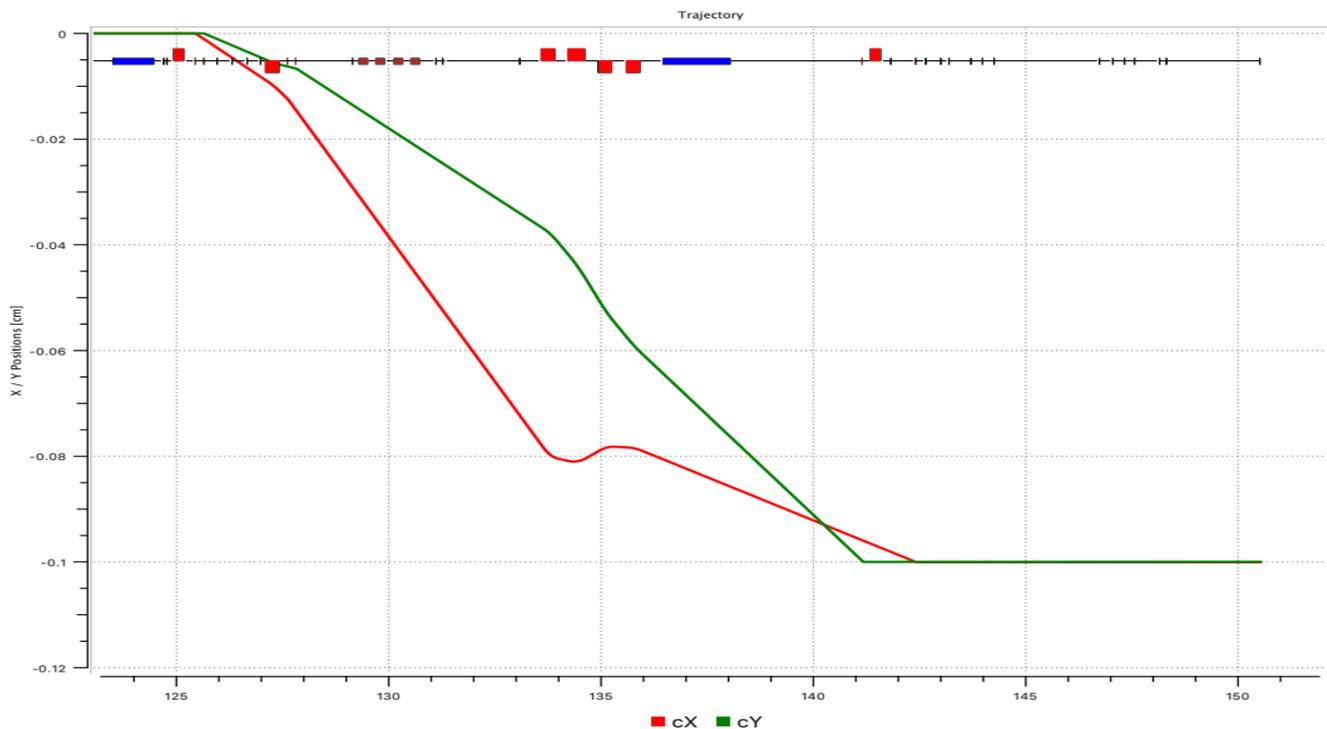

**Figure 9**. Correction of 1 mm target offsets in both X and Y in the interim optics solution.

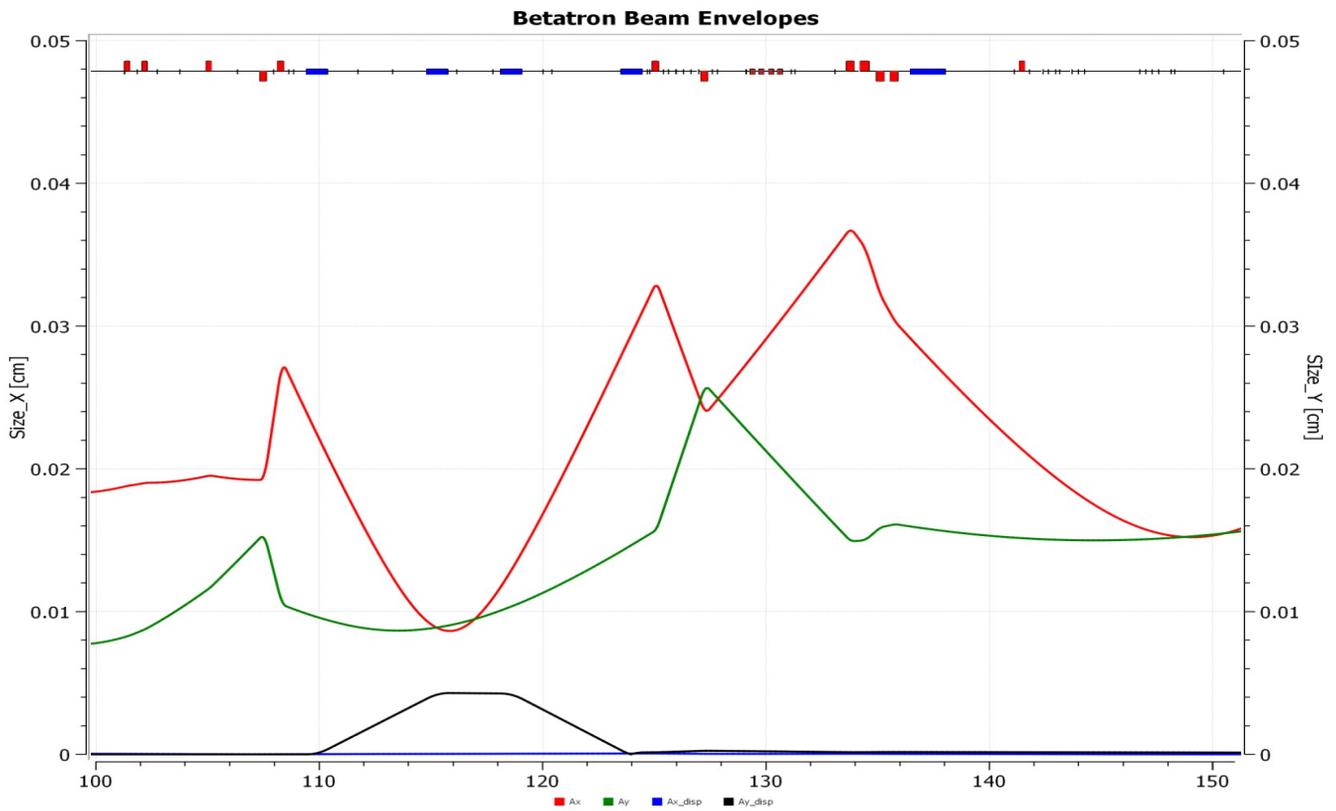

**Figure 10**. Beam envelopes of interim layout optimized for spot size at pivot (right edge of plot)

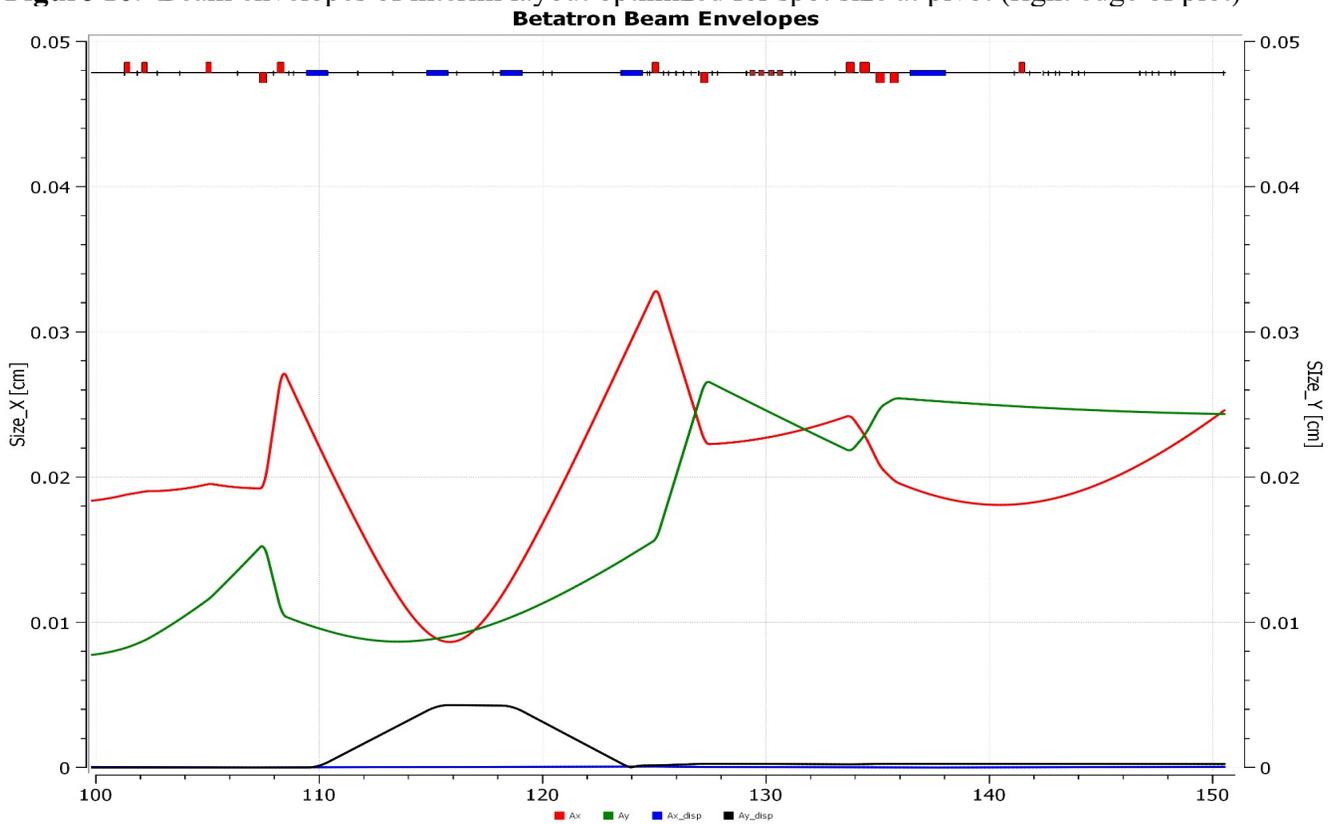

**Figure 11**. Beam envelopes of interim layout optimized for raster at pivot (right edge of plot). While the spot size is two-thirds larger than in Figure 10, it is still below specification (0.03 cm, 300 microns)

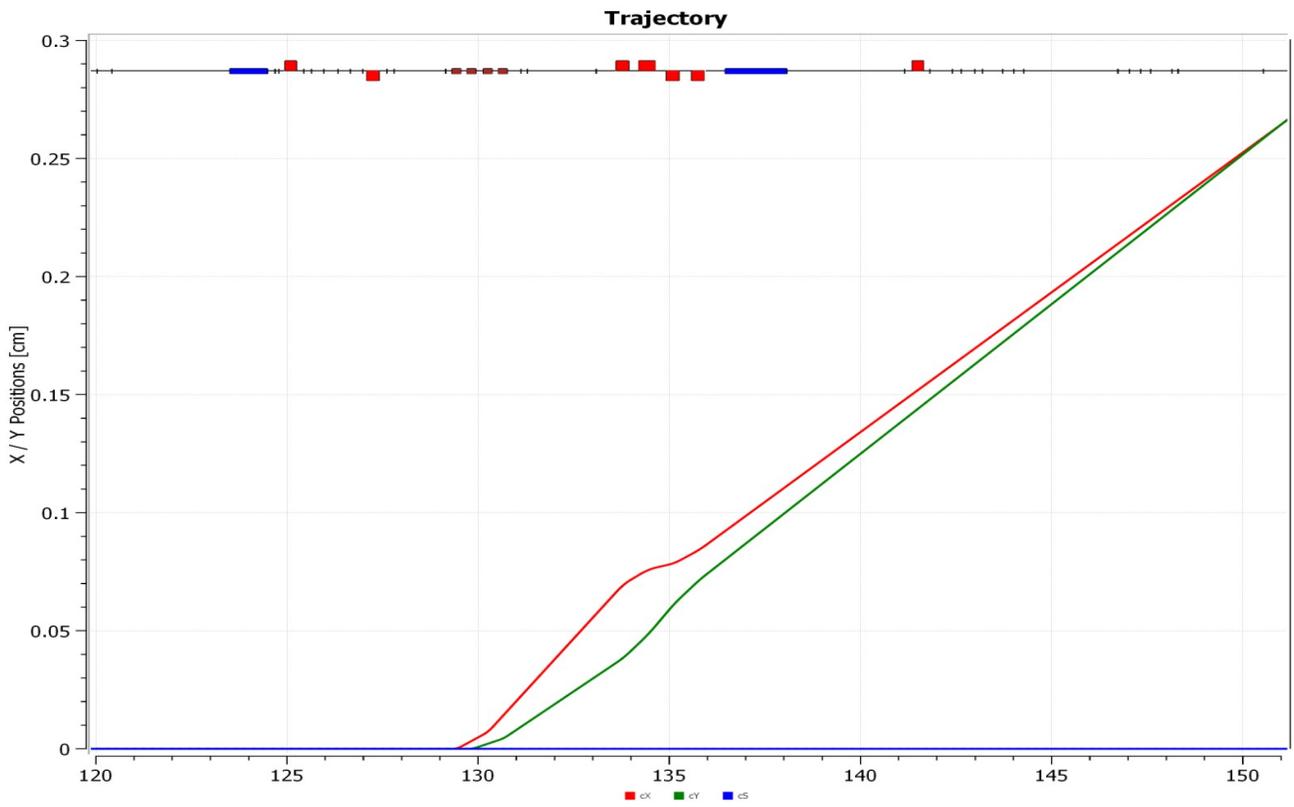

**Figure 12**. Raster with optics optimized for spot size at pivot, Figure 10.

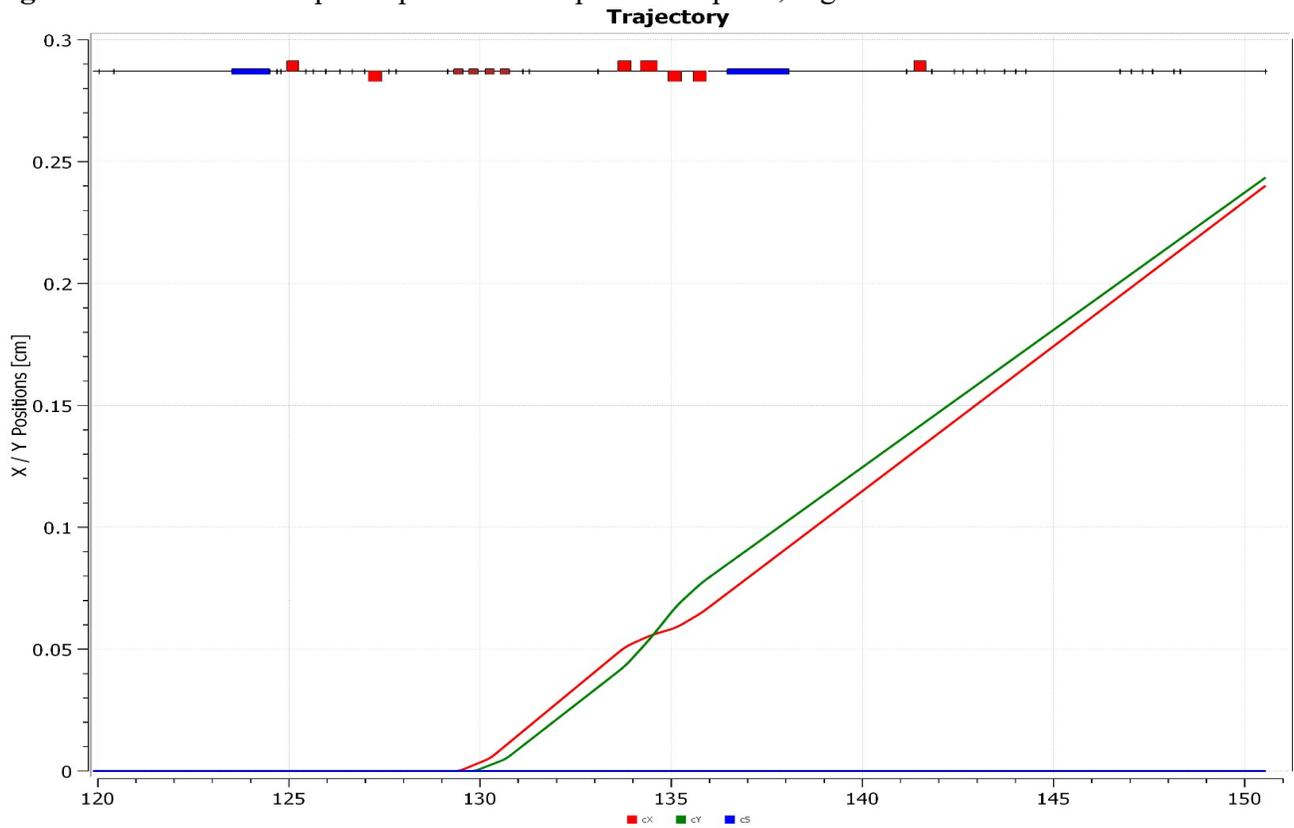

**Figure 13** Raster with optics optimized for near-equal coil currents, so a circular raster would be easy to implement. Envelopes in Figure 11.

This interim solution will be re-used if the SoLID suite of experiments [14] is funded to follow MOLLER. One of the approved experiments uses a polarized frozen ammonia target which requires 2 cm diameter raster pattern to minimize depolarization by the beam. This is not possible to obtain with just one pair of raster coils even with a target 9 m downstream of the MOLLER target, 4.5 m downstream of the Pivot. Two pair of coils will suffice.

**Conclusions**

A beam line based on two decades of experience with parity experiments has been designed. The first half of the line should be installed in the first part of 2022 to gain experience with the altered Moller polarimeter and provide better beam to the experiments scheduled for that year. The remainder will be installed with the MOLLER experiment and will become the permanent Hall A beam line.

**Acknowledgements**

This material is based upon work supported by the U.S. Department of Energy, Office of Science, Office of Nuclear Physics under contract DE-AC05-06OR23177.
Roger Carlini participated in numerous discussions with the first author about this beam line. Greg Smith provided valuable suggestions on the text.

**16** Yuan Mei, Shujie Li, John Arrington, Joseph Camilleri, Aled Cuda, James Egelhoff, Yury Kolomensky, Ernst Sichtermann "A direct-sampling RF receiver for MOLLER beam charge measurement" https://arxiv.org/abs/2110.09575 submitted to JINST

**Appendix: Beam line drawings**

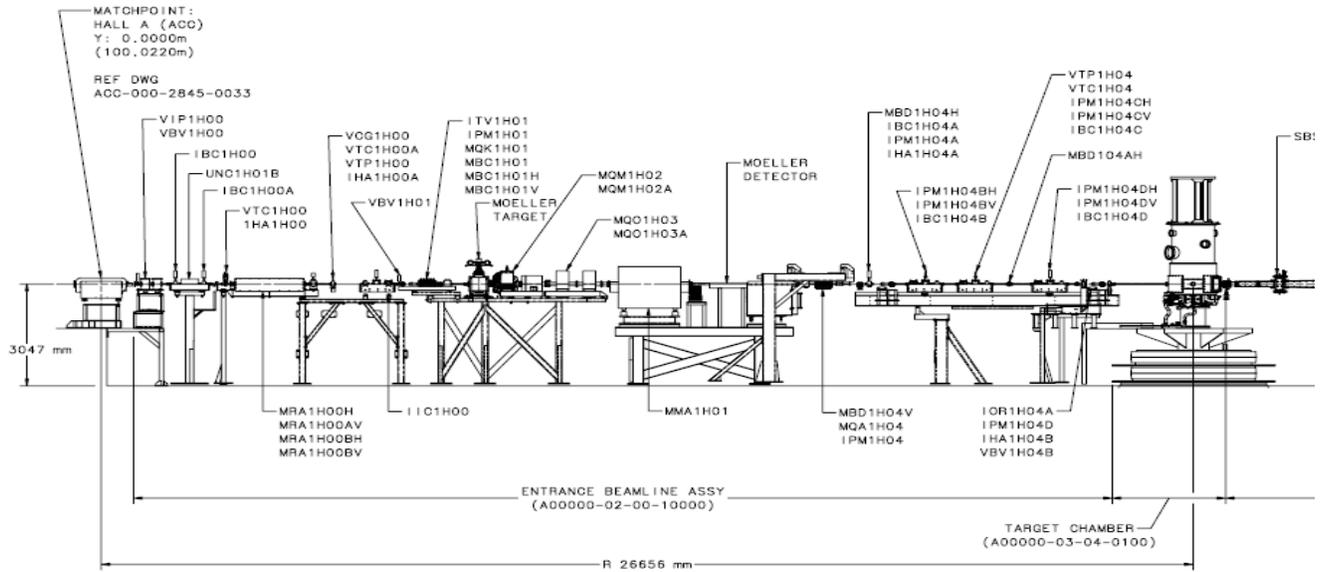

**Figure A1**. Existing beam line

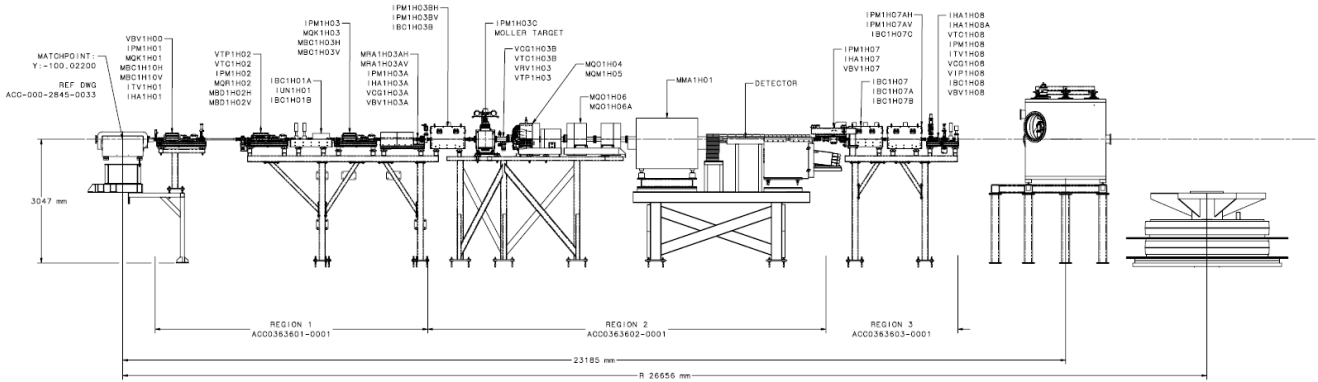

**Figure A2**. MOLLER beam line